\documentclass{article}
\usepackage{ijcai24}

\usepackage{times}
\usepackage{soul}
\usepackage{url}
\usepackage[hidelinks]{hyperref}
\usepackage[utf8]{inputenc}
\usepackage[small]{caption}
\usepackage{graphicx}
\usepackage{amsmath}
\usepackage{amsthm}
\usepackage{booktabs}
\usepackage{algorithm}
\usepackage{algorithmic}
\usepackage[switch]{lineno}

\usepackage{inconsolata}
\usepackage{listings}
\usepackage{graphicx}
\usepackage{comment}
\usepackage{xcolor}

\definecolor{codegreen}{rgb}{0,0.6,0}
\definecolor{codegray}{rgb}{0.5,0.5,0.5}
\definecolor{codepurple}{rgb}{0.58,0,0.82}
\definecolor{backcolour}{rgb}{0.95,0.95,0.92}

\lstdefinestyle{mystyle}{
    backgroundcolor=\color{backcolour},   
    commentstyle=\color{codegreen},
    keywordstyle=\color{magenta},
    numberstyle=\tiny\color{codegray},
    stringstyle=\color{codepurple},
    basicstyle=\ttfamily\scriptsize,
    breakatwhitespace=false,         
    breaklines=true,                 
    captionpos=b,                    
    keepspaces=true,                 
    numbers=left,                    
    numbersep=5pt,                  
    showspaces=false,                
    showstringspaces=false,
    showtabs=false,                  
    tabsize=2
}

\title{CodeMirage: Hallucinations in Code Generated by Large Language Models}

\author{
Vibhor Agarwal$^1$\thanks{Work done during internship at JP Morgan AI Research.}
\and
Yulong Pei$^2$\and
Salwa Alamir$^2$\And
Xiaomo Liu$^3$\\
\affiliations
$^1$University of Surrey, Surrey, UK \\
$^2$JP Morgan AI Research, London, UK \\
$^3$JP Morgan AI Research, New York, USA \\
\emails
v.agarwal@surrey.ac.uk,
\{yulong.pei, salwa.alamir, xiaomo.liu\}@jpmorgan.com
}

\begin{document}

\maketitle

\begin{abstract}
Large Language Models (LLMs) have shown promising potentials in program generation and no-code automation. However, LLMs are prone to generate hallucinations, i.e., they generate text which sounds plausible but is incorrect. Although there has been a recent surge in research on LLM hallucinations for text generation, similar hallucination phenomenon can happen in code generation. Sometimes the generated code can have syntactical or logical errors as well as more advanced issues like security vulnerabilities, memory leaks, etc. Given the wide adaptation of LLMs to enhance efficiency in code generation and development in general, it becomes imperative to investigate hallucinations in code generation. To the best of our knowledge, this is the first attempt at studying hallucinations in the code generated by LLMs. We start by introducing the code hallucination definition and a comprehensive taxonomy of code hallucination types. We propose the first benchmark \textbf{CodeMirage} dataset\footnote{The dataset will be released upon acceptance.} for code hallucinations. The benchmark contains $1,137$ GPT-3.5 generated hallucinated code snippets for Python programming problems from two base datasets – \textit{HumanEval} and \textit{MBPP}. We then propose the methodology for code hallucination detection and experiment with open source LLMs such as CodeLLaMA as well as OpenAI's GPT-3.5 and GPT-4 models using one-shot prompt. We find that GPT-4 performs the best on HumanEval dataset and gives comparable results to the fine-tuned CodeBERT baseline on MBPP dataset. Towards the end, we discuss various mitigation strategies for code hallucinations and conclude our work. 
\end{abstract}

\section{Introduction}

\begin{figure}
    \centering
\begin{lstlisting}[language=Python]
import yaml

def yaml_load(filename):
    """Open a file, load its contents into a variable with yaml, and return it."""
    with open(filename, 'f') as f:
        data = yaml.load(f, Loader=yaml.Baseloader)
        return data
\end{lstlisting}
    \caption{An example of robustness issue in the LLM generated code. It will raise ``FileNotFoundException'' in case the file does not exist.}
    \label{fig:robustness-issue-eg}
\end{figure}

Large Language Models (LLMs) have shown great capabilities pushing forward the field of artificial intelligence in general and natural language generation in particular. OpenAI's GPT-3.5~\cite{ouyang2022training} and GPT-4~\cite{openai2023gpt4} as well as open source LLMs such as LLaMA-3.1~\cite{llama3.1}, LLaMA-2~\cite{touvron2023llama}, etc. have demonstrated strong performance in various natural language processing (NLP) tasks~\cite{yang2024harnessing,zhu2023can,agarwal2023haterephrase}. LLMs have also been trained on large snippets of programming codes and have shown strong performance in the field of code generation. GPT-3.5 and GPT-4 as well as open source models such as CodeLLaMA~\cite{roziere2023code} can generate code in various programming languages.

Albeit LLMs show remarkable capabilities, they frequently hallucinate, i.e., they generate text that sounds plausible but is incorrect. This makes text hallucination detection task very challenging as the generated text is often similar in style but is factually incorrect or conflicting with the input or the context~\cite{zhang2023siren}. According to \cite{ji2023survey}, hallucination is the \textit{generated content that is nonsensical or unfaithful to the provided source content}. Similar hallucination phenomenon can happen in the code generated by LLMs as well. Sometimes the generated code can have syntactical and logical errors as well as more advanced issues like security vulnerabilities, memory leaks, robustness issues, etc. Figure~\ref{fig:robustness-issue-eg} shows an example of robustness issue in the LLM generated code snippet. LLM generates a function to load YAML file but does not check if the file exists before reading it. Therefore, the generated code snippet has robustness issue since it will raise ``FileNotFoundException'' in case the file does not exist but does not handle the exception adequately. Considering the adaptation of LLMs in industrial environment to help code generation and development is becoming more widespread~\cite{yang2023harnessing}, it is very important to detect and mitigate code hallucinations.

For the first time, we study hallucinations in the code generated by LLMs. We firstly introduce the code hallucination definition and a comprehensive taxonomy of code defects that LLMs can hallucinate. We then propose the first benchmark dataset -- \textit{CodeMirage} containing GPT-3.5 generated hallucinated Python code snippets and then experiment and measure LLM capabilities for automatically detecting code hallucinations using one-shot prompts. We believe that this work, including the comprehensive taxonomy, new dataset and insights, can open new avenues for research and development in both academia and industry.

The contributions of our paper are summarized as follows:
\begin{itemize}
    \item To the best of our knowledge, we introduce the problem of code hallucination for the first time and provide its definition and a comprehensive taxonomy of code defects that LLMs can hallucinate.
    \item We introduce the first benchmark dataset -- \textit{CodeMirage} containing GPT-3.5 generated $1,137$ Python code snippets with $5$ hallucination types.
    \item For code hallucination detection, we introduce several baselines and experiment with open source LLMs as well as OpenAI's GPT models. Although code hallucination detection is a challenging task, LLMs demonstrate reasonable performance in detecting various hallucination types.
    \item We conduct comprehensive experiments and ablation studies to demonstrate the capabilities of LLMs for code hallucination detection and discuss various code hallucination mitigation strategies in the future work. 
\end{itemize}

\section{Related Work}\label{sec:related-work}

\subsection{Language Models for Code Generation}


The triumph of language models in natural language modeling has brought interest among researchers and practitioners on using language models for code generation. Code generation refers to generating programs that need to satisfy all the constraints defined by the underlying task such as test cases, problem descriptions, etc. Pre-trained transformer-based models such as CodeBERT~\cite{feng2020codebert} are specifically trained for code generation using Masked Language Modeling and Replaced Token Detection training objectives. Decoder pre-trained models are designed to predict the next token based on a given input context. OpenAI's GPT-series~\cite{radford2018improving} are decoder-based models for text generation. Based on GPT-2, \cite{lu2021codexglue} released CodeGPT for code completion and text-to-code generation tasks. After GPT-3 was developed, CodeX\footnote{\url{https://openai.com/blog/openai-codex}, last accessed 23 May 2024.} and GitHub Copilot\footnote{\url{https://github.com/features/copilot}, last accessed 23 May 2024.} were released for code generation. After the success of ChatGPT, OpenAI's GPT-3.5~\cite{ouyang2022training} and GPT-4~\cite{openai2023gpt4} models became state-of-the-art for natural language generation. They have shown good performance in code generation as well~\cite{poldrack2023ai}, but these models are proprietary. Similar open source models such as LLaMA-2~\cite{touvron2023llama} and CodeLLaMA~\cite{roziere2023code} were released for natural language generation tasks. CodeLLaMA is an open sourced LLM based on LLaMA-2 itself but fine-tuned for code-related tasks such as code generation. Another relevant direction is detecting software vulnerability using LLMs where different LLMs~\cite{jensen2024software} and other information such as code structures~\cite{lu2024grace} have been explored to detect vulnerabilities in the code.

\subsection{Hallucinations in LLMs}

Although LLMs have shown remarkable performance in natural language generation, they still inevitably encounter several issues, hallucination being one of the top~\cite{wu2024new,ghafouri2023ai,huang2023bias}. \cite{ji2023survey} defined hallucination as the generated content that is nonsensical or unfaithful to the provided source content. Previous works~\cite{ji2023survey,maynez2020faithfulness,huang2021factual} categorized hallucination into two main categories -- \textit{intrinsic} and \textit{extrinsic}. Intrinsic hallucination happens when the generated output contradicts the source content, whereas extrinsic hallucination happens when the generated output cannot be verified from the source content (i.e., output that can neither be supported nor contradicted by the source). Within the context of LLMs, \cite{zhang2023siren} defined hallucination into 3 categories -- \textit{input-conflicting} (LLM-generated content deviates from the source input provided by users), \textit{context-conflicting} (LLM-generated content conflicts with previously generated information by itself), and \textit{fact-conflicting} (LLM-generated content is not faithful to the established world knowledge) hallucinations.

Similar to text hallucinations, LLMs can hallucinate during code generation as well. Sometimes the LLM-generated code can have syntactical or logical errors as well as more advanced issues like security vulnerabilities, memory leaks, etc. There is no prior work that specifically look at hallucinations in the generated code. \cite{dinh2023large} studied the buggy-code completion problem
and find that the presence of potential bugs in the code context significantly degrades the code generation performance of the LLMs. \cite{liu2023your} evaluated the functional correctness of the LLM-generated code with large amounts of test-cases newly produced by an automatic test input generator.
\cite{ouyang2023llm} conducted empirical study to measure non-determinism in code generated by LLMs. They find that results from LLMs can be highly unstable; non-deterministically returning very different codes for the same prompt. \cite{bhatt2023purple} introduced CyberSecEval, a benchmark developed to help bolster the cybersecurity of LLMs employed as coding assistants. They find a high tendency of more advanced models to suggest insecure code, highlighting the critical need for integrating security considerations in the development of sophisticated LLMs.

For the first time in the literature, we study hallucinations in the LLM-generated code. We introduce the code hallucination definition, a comprehensive taxonomy and then propose the first benchmark dataset -- \textit{CodeMirage} containing GPT-3.5 generated hallucinated Python code snippets followed by the methodology for detecting code hallucinations.

\begin{table*}
    \centering
    \begin{tabular}{p{17cm}}
    \hline
        \textit{I want you to act as a code hallucination generator. Given the function signature, task description, and test cases, your objective is to write the hallucinated python code that sounds plausible but is incorrect by \texttt{<specific code hallucination type description>}. Below is an example:}
    
        \textit{\texttt{<Example code hallucination>}}
        
        \textit{You should try your best to generate a hallucinated code to the following question:}
        
        \textit{\texttt{<Programming question>}} \\
    \hline
    \end{tabular}
    \caption{Layout of Code Hallucination Generation Prompt.}
    \label{tab:hallu-gen-prompt}
\end{table*}

\section{Hallucinations in Code Generation}\label{sec:code-hallu-taxonomy}

In this section, we formally introduce our problem statement of code hallucination in Section~\ref{sec:problem-def} and then present a comprehensive taxonomy of \textit{five} hallucination types that can occur in code generated by LLMs in Section~\ref{sec:taxonomy}.

\subsection{Problem Definition}\label{sec:problem-def}

Large Language Models have shown good performance in code generation. However, sometimes the generated code may sound plausible but can have several code defects such as security vulnerabilities. We define \textit{\textbf{hallucinated code} as the generated code that has one or more code defects such as dead or unreachable code, syntactic or logical errors, robustness issues such as the code fails on edge cases or raises an exception, or has security vulnerabilities or memory leaks}.

\subsection{Taxonomy}\label{sec:taxonomy}

Based on different types of code defects that can occur, we define the following \textit{five} hallucination categories for the code generated by LLMs.

\begin{itemize}
    \item \textbf{\textit{Dead or Unreachable code}}: Generated code has dead, unreachable or redundant piece of code.
    \item \textbf{\textit{Syntactic incorrectness}}: Generated code has syntactic errors and therefore, fails to compile.
    \item \textbf{\textit{Logical error}}: Generated code has logical errors, i.e., the generated code cannot solve the given problem correctly.
    \item \textbf{\textit{Robustness issue}}: Generated code has robustness issues such as it fails on certain edge cases or raises an exception (does not perform required exception handling).
    \item \textbf{\textit{Security vulnerabilities}}: Generated code has security vulnerabilities or memory leaks.
\end{itemize}

\section{CodeMirage Dataset}\label{sec:CodeMirage-dataset}

In this section, we discuss the details of our CodeMirage dataset. We begin with describing the dataset generation methodology (Section~\ref{sec:data-generation}) followed by verifying the generated dataset via human annotations (Section~\ref{sec:human-annotations}). Then we share dataset statistics and various characteristics of the dataset in Section~\ref{sec:dataset-stats}.

\subsection{Dataset Generation}\label{sec:data-generation}

For generating the code hallucination dataset -- \textbf{CodeMirage}, we select two popular base datasets -- HumanEval~\cite{chen2021evaluating} and MBPP~\cite{austin2021program}. HumanEval dataset~\cite{chen2021evaluating} contains $164$ Python programming problems with function signatures, problem description as docstrings, programming solution and test cases for evaluation. Similarly, MBPP~\cite{austin2021program} benchmark consists of $973$ crowd-sourced Python programming problems, designed to be solvable by entry-level programmers, covering programming fundamentals, standard library functionality, and so on. Each problem consists of a task description, code solution and 3 automated test cases.

For generating hallucinated code snippets, we use GPT-3.5. We design explicit prompts for each of the hallucination types and input them into GPT-3.5 model to get Python code generations that have specific hallucination types. Table~\ref{tab:hallu-gen-prompt} shows the layout of code hallucination generation prompt. Each prompt has code hallucination definition describing a specific type of hallucination and an example showcasing a sample problem statement along with the hallucinated code and test cases. Specific prompts for different hallucination types are mentioned in Appendix~\ref{sec:appendix1}. For every problem in both the datasets, we randomly select one of the five hallucination types and then input type specific prompt along with the problem description and test cases. As a result, we get hallucinated code as an output from GPT-3.5 model and we assign the selected hallucination type as the gold label, further validated through human annotations as described in Section~\ref{sec:human-annotations}.

\subsection{Human Annotations}\label{sec:human-annotations}

To validate the dataset and type specific hallucinated code generations, we conduct human evaluation. We randomly selected $200$ programming problems and solutions from both the datasets ($50$ from HumanEval and $150$ from MBPP) in proportion to the number of problems in each of the datasets. We selected a balanced sample with all the hallucination types in equal numbers. We give detailed instructions of the annotation task, definitions for each of the code hallucination types and an example for each. We then ask the annotators to annotate each Python code snippet as one of the \textit{five} code hallucination types: ``dead code'', ``syntactic incorrectness'', ``logical error'', ``robustness issue'', ``security vulnerabilities'' or ``no hallucination''. Annotations were performed by $5$ human annotators, well-versed in Python programming, with every annotator annotating about $100$ Python code snippets. Initially, each code snippet was annotated by $2$ annotators. In case of label conflicts, we introduced a third annotation. Therefore, every code snippet in the sampled dataset has at least $2$ annotations and in case of conflicts, $3$ annotations so that we have a majority label. Overall, annotators get an average Cohen's kappa score of $0.76$ which denotes strong agreement. We also measure accuracy between the annotated majority labels and the automatic gold labels we create during the dataset generation process. High accuracy of $0.81$ denotes that GPT-3.5 has generated code snippets as per the specific hallucination type and that automatically assigned gold labels for hallucination types are reliable.

\begin{table}
    \centering
    \small
    \begin{tabular}{l|c|c|c}
    \hline
    \textbf{Hallucination Type} & \textbf{HumanEval} & \textbf{MBPP} & \textbf{Total}  \\
    \hline
    Dead/unreachable code & 36 & 190 & 226   \\
    Syntactic incorrectness & 38 & 214 & 252   \\
    Logical error & 31 & 189 & 220 \\
    Robustness issue & 26 & 170 & 196 \\
    Security vulnerabilities & 33 & 210 & 243 \\
    \hline
    \textbf{Total} & \textbf{164} &  \textbf{973} &  \textbf{1137} \\
    \hline
    \end{tabular}
    \caption{\textit{CodeMirage} dataset statistics.}
    \label{tab:data-stats}
\end{table}
\begin{table}
    \centering
    \small
    \begin{tabular}{l|c|c}
    \hline
    \textbf{Hallucination Type} & \textbf{HumanEval} & \textbf{MBPP}  \\
    \hline
    Ground-truth correct code & 2.98 & 2.27  \\
    \hline
    Dead/unreachable code & 4.42 & 4.17   \\
    Syntactic incorrectness & N/A & N/A   \\
    Logical error & 3.71 & 3.11 \\
    Robustness issue & 3.19 & 2.77 \\
    Security vulnerabilities & 4.27 & 3.17 \\
    \hline
    \end{tabular}
    \caption{McCabe Cyclomatic Complexity of \textit{CodeMirage} dataset.}
    \label{tab:mccabe-complexity}
\end{table}

\begin{table*}
    \centering
    \begin{tabular}{p{17cm}}
    \hline
        \textit{I want you to act as a code judge. Given the task description, function signature, and the generated code, your objective is to detect if the generated code has defects, incorrect code or hallucinations. Hallucinated code is the generated code that has one or more code defects such as dead or unreachable code, syntactic or logical errors, robustness issues such that the code fails on certain test cases or raises an exception, or has security vulnerabilities or memory leaks. Below are the 5 categories for code hallucination along with an example:}

        \textit{1. Dead code: Generated code has dead, unreachable or redundant piece of code.
        Example:}
        
        \textit{\texttt{<Example code snippet>}}
        
        \textit{2. Syntactic incorrectness: Generated code has syntactic errors and therefore, fails to compile.
        Example:}
        
        \textit{\texttt{<Example code snippet>}}
        
        \textit{3. Logical error: Generated code has logical errors.
        Example:}
        
        \textit{\texttt{<Example code snippet>}}
        
        \textit{4. Robustness issue: Generated code has robustness issues as it fails on certain test cases or raises an exception due to poor exception handling.
        Example:}
        
        \textit{\texttt{<Example code snippet>}}
        
        \textit{5. Security vulnerabilities: Generated code has security vulnerabilities or memory leaks. Example:}
        
        \textit{\texttt{<Example code snippet>}}
        
        \textit{You should try your best to determine if the code contains any defects or hallucinations based on the above hallucination types. Please output only 1 out of these 6 classes: [``no hallucination'', ``dead code'', ``syntactic incorrectness'', ``logical error'', ``robustness issue'', ``security vulnerabilities''].}
        
        \textit{\texttt{<Programming question and code snippet>}}    \\
    \hline
    \end{tabular}
    \caption{Layout of Code Hallucination Detection Prompt.}
    \label{tab:hallu-det-prompt}
\end{table*}

\subsection{Dataset Statistics}\label{sec:dataset-stats}

Table~\ref{tab:data-stats} shows the number of hallucinated Python code snippets in \textit{CodeMirage} dataset for each of the 5 code hallucination types with individual splits for base datasets -- \textit{HumanEval} and \textit{MBPP}. In total, \textit{CodeMirage} dataset has $1,137$ programming problems, LLM-generated hallucinated Python code snippets, ground truth code snippets, and test cases to evaluate code snippets.


To measure the complexity of generated code snippets, we compute McCabe's cyclomatic complexity~\cite{mccabe1976complexity}. Cyclomatic complexity is a quantitative measure to indicate the complexity of a program by measuring the number of linearly independent paths through a program’s source code. Table~\ref{tab:mccabe-complexity} shows average cyclomatic complexity scores for CodeMirage dataset for each of the base datasets. We compute and compare complexity scores of ground-truth code snippets with the hallucinated code snippets. Cyclomatic complexity scores for ``syntactic incorrectness'' hallucination type can not be computed due to syntax errors (denoted by N/A in Table~\ref{tab:mccabe-complexity}). For all other hallucination types, average cyclomatic complexity scores are higher than ground truth code snippets. As expected, code snippets with dead code as hallucinations have the highest cyclomatic complexity. On the other hand, ``robustness issue'' hallucination type has the lowest cyclomatic complexity for both HumanEval and MBPP datasets since it is expected for the code snippet to be robust and hence, lower complexity even though it contains minor robustness issues. This consistent behavior demonstrates the effectiveness of our approach for generating the code hallucination dataset.

\section{Code Hallucination Detection}\label{sec:code-hallu-detect}
Detecting code hallucinations is a challenging task as the code snippet may seem to be plausible but can be incorrect as it can have issues such as security vulnerabilities, memory leaks, etc. which are often hard to detect. After describing the \textit{CodeMirage} dataset in the previous section, we discuss the methodology and results for code hallucination detection in this section.

\subsection{Methodology}
For code hallucination detection, we prompt various large language models to detect whether a code snippet has hallucination and if present, the type of hallucination. We develop an one-shot prompt asking LLMs to detect five hallucination types as well as ``no hallucination'' category, given the problem description and code snippet as mentioned in Appendix~\ref{sec:appendix2}. We also provide the definitions and an example for each type of hallucinations in the prompt. Table~\ref{tab:hallu-det-prompt} shows the layout of code hallucination detection prompt. We experiment with $3$ LLMs -- an open source CodeLLaMA model as well as OpenAI's GPT-3.5 and GPT-4 models for detecting code hallucinations. We describe various LLMs and baselines used for detecting code hallucinations below:

\begin{itemize}
    \item \textbf{CodeLLaMA}: CodeLLaMA~\cite{roziere2023code} is an open source LLM for code based on LLaMA-2~\cite{touvron2023llama} providing state-of-the-art performance among open models, infilling capabilities, support for large input contexts, and zero-shot instruction following ability for programming tasks. We use CodeLLaMA-7B-Instruct model having 7 billion parameters and fine-tuned to follow instructions.
    
    \item \textbf{GPT-3.5}: We use OpenAI's GPT-3.5~\cite{ouyang2022training} model, accessed through OpenAI's official API.
    
    \item \textbf{GPT-4}: We also experiment with GPT-4~\cite{openai2023gpt4}, the OpenAI's state-of-the-art model, accessed through its official API.
    
    \item \textbf{CodeBERT}: CodeBERT~\cite{feng2020codebert} is a pre-trained transformer-based model for programming language, which is a multi-programming-lingual model pre-trained in $6$ programming languages. As a baseline, we fine-tune CodeBERT on our CodeMirage dataset with 80:20 split for training and testing sets, respectively. For the train set, we follow stratified sampling for each of the two base datasets. We keep test sets separate to evaluate the performance of fine-tuned model separately on both the base datasets.
\end{itemize}

\subsection{Experimental Setup and Evaluation Metrics}

For detecting code hallucinations, we experiment with one-shot prompt and input it into LLMs along with problem description and code snippet. For CodeLLaMA-Instruct, we use its open-source implementation after downloading the model weights of $7$ billion parameters. For OpenAI's GPT-3.5 and GPT-4 models, we use their official API\footnote{\url{https://openai.com/product}, last accessed 28 Dec 2023.}. We set a temperature of $0.7$ and maximum number of tokens for generation to $256$.

\noindent \textbf{Evaluation Metrics}. Since we model code hallucination detection task as a multi-class classification task predicting either of the $5$ code hallucination types or ``no hallucination'' category, we use accuracy, macro-precision, macro-recall, and macro-F1 scores for performance evaluation.

\subsection{Results}

In this section, we discuss results of various language models for detecting code hallucinations. Table~\ref{tab:results} shows performance scores for various LLMs for code hallucination detection on CodeMirage dataset. CodeBERT, fine-tuned on our CodeMirage dataset, achieves an accuracy of $0.5938$ and macro-F1 score of $0.4897$ on HumanEval dataset, whereas it achieves an accuracy of $0.6825$ and macro-F1 score of $0.6344$ on MBPP dataset. CodeLLaMA, an open source LLM, does not perform well when prompted for code hallucination detection as it achieves macro-F1 scores of $0.0424$ and $0.0271$ on HumanEval and MBPP datasets, respectively. Surprisingly, there is a big performance gap between GPT-3.5 and GPT-4 models for code hallucination detection with the same prompt.
On one hand, GPT-3.5 achieves macro-F1 scores of $0.2654$ and $0.2092$ on HumanEval and MBPP, respectively. On the other hand, GPT-4 model achieves the best performance with an overall macro-F1 score of $0.5512$ for HumanEval and second best score of $0.5195$ for MBPP. From Table~\ref{tab:results}, we can infer that GPT-4 model with just one-shot prompt performs the best on HumanEval dataset and beats the fine-tuned CodeBERT model by $6.15$ percentage macro-F1 score. On the other hand, GPT-4 model gives second best performance on MBPP dataset in terms of macro-F1 score and could not beat the fine-tuned CodeBERT model but shows comparable results. However, GPT-4 still gives the best macro-Precision score. Overall, we can conclude that LLMs, especially GPT-4, performs comparable, if not better, with fine-tuned CodeBERT model with mere one-shot prompt for code hallucination detection.

\begin{table*}
    \centering
    \begin{tabular}{l|cccc|cccc}
    \hline
      & \multicolumn{4}{c|}{\textbf{HumanEval}} & \multicolumn{4}{c}{\textbf{MBPP}}  \\
    \hline
    \textbf{LLM} & \textbf{Acc.} & \textbf{ma-P} & \textbf{ma-R} & \textbf{ma-F1} & \textbf{Acc.} & \textbf{ma-P} & \textbf{ma-R} & \textbf{ma-F1} \\
    \hline
    CodeBERT & \textbf{0.5938} & \underline{0.5547} & \underline{0.5162} & \underline{0.4897} & \textbf{0.6825} & \underline{0.6455} & \textbf{0.6598} & \textbf{0.6344}   \\ 
    CodeLLaMA & 0.0250 & 0.2800 & 0.0232 & 0.0424 & 0.0146 & 0.4061 & 0.0145 & 0.0271   \\
    GPT-3.5 & 0.2134 & 0.5552 & 0.1956 & 0.2654 & 0.1614 & 0.5549 & 0.1530 & 0.2092   \\
    GPT-4   & \underline{0.5915} & \textbf{0.6710} & \textbf{0.5514} & \textbf{0.5512} & \underline{0.5735} & \textbf{0.6644} & \underline{0.5626} & \underline{0.5195}    \\
    \hline
    \end{tabular}
    \caption{Results for code hallucination detection on \textit{CodeMirage} dataset. Best scores are in \textbf{bold} and second best scores are \underline{underlined}.}
    \label{tab:results}
\end{table*}

\section{Conclusions and Future Work}\label{sec:conclusions}

LLMs have shown good performance in code generation. In this work, we study hallucinations for the first time in the code generated by LLMs. At first, we introduce the code hallucination definition and a comprehensive taxonomy of 5 hallucination types. We then propose the first ever CodeMirage dataset containing $1,137$ GPT-3.5 generated hallucinated Python code snippets. We believe this comprehensive code hallucination taxonomy and the new dataset can open new avenues for research and development in both academia and industry to evaluate code snippets generated by LLMs and mitigate code defects. We also experiment with various open source as well as OpenAI's GPT-like LLMs for detecting code hallucinations using one-shot prompts. We find that GPT-4 model performs the best on HumanEval dataset, while it performs second best on MBPP dataset.

In general, we conclude that LLMs, especially GPT-4, performs comparable, if not better, with fine-tuned CodeBERT model for code hallucination detection. The overall performance suggests that the task of code hallucination detection and the CodeMirage dataset are challenging as even the state-of-the-art fine-tuned transformer-based models and LLMs can not performance well with high macro-F1 score. As a result, there is a huge scope for future works. For code hallucination detection, fine-tuning LLMs~\cite{hu2023llm} with specific hallucination detection instructions can improve the performance. We can also use software engineering methods such as using compilers, abstract syntax trees (ASTs)~\cite{shippey2015exploiting}, control flow graphs (CFGs)~\cite{phan2017convolutional,zhang2023detecting}, etc. to detect code defects. Another important direction of research can be mitigating code hallucinations. We can use several mitigation strategies as also used in text hallucinations~\cite{ji2023survey,zhang2023siren} such as knowledge-enhanced prompt tuning~\cite{ma2023codeprompt}, retrieval-augmented code generation~\cite{parvez2021retrieval}, fine-tuning LLMs~\cite{hu2023llm}, etc. to mitigate code hallucinations. Similarly, software engineering techniques analyzing execution workflows of the generated code snippets can be used to detect and mitigate code defects. Therefore, there are a lot of opportunities for future work to effectively detect and mitigate code hallucinations. Solving this problem is a huge step forward for companies to be able to safely adopt the use of LLMs for code generation in practice.

\section*{Disclaimer}
This paper was prepared for informational purposes by the Artificial Intelligence Research group of JPMorgan Chase \& Co. and its affiliates (``JP Morgan'') and is not a product of the Research Department of JP Morgan. JP Morgan makes no representation and warranty whatsoever and disclaims all liability, for the completeness, accuracy or reliability of the information contained herein. This document is not intended as investment research or investment advice, or a recommendation, offer or solicitation for the purchase or sale of any security, financial instrument, financial product or service, or to be used in any way for evaluating the merits of participating in any transaction, and shall not constitute a solicitation under any jurisdiction or to any person, if such solicitation under such jurisdiction or to such person would be unlawful.

\bibliographystyle{named}
\bibliography{ijcai24}

\appendix

\section{Prompts for Code Hallucination Generation}\label{sec:appendix1}

\textbf{Dead/Unreachable code}: \textit{I want you to act as a code hallucination generator. Given the function signature, task description, and test cases, your objective is to write the hallucinated python code that sounds plausible but is incorrect by inserting a dead or unreachable or redundant code. Below is an example:}

\begin{verbatim}
from typing import List

def has_close_elements(numbers: List[float],
    threshold: float) -> bool:
    """Check if in given list of
    numbers, are any two numbers closer 
    to each other than given threshold.
    """
>>> has_close_elements(
    [1.0, 2.0, 3.0], 0.5)
False
>>> has_close_elements(
    [1.0, 2.8, 3.0, 4.0, 5.0, 2.0], 0.3)
True

def has_close_elements(
    numbers: List[float], threshold: float)
    -> bool:
    if threshold < 0:
        return False
        
    numbers.sort()
    dead_code = [i for i in numbers if i < 0]
    
    for i in range(len(numbers) - 1):
        if abs(numbers[i] - numbers[i + 1])
            < threshold:
            return True
        else:
            dead_code.append(numbers[i])
    dead_code = sorted(dead_code, 
                reverse=True)
    return False
\end{verbatim}

\textit{You should try your best to generate a hallucinated code to the following question:}
\texttt{<Programming question>}

\textbf{Syntactic incorrectness}: \textit{I want you to act as a code hallucination generator. Given the function signature, task description, and test cases, your objective is to write the hallucinated python code that sounds plausible but is incorrect by making syntax errors in the code. Below is an example:}

\begin{verbatim}
from typing import List

def has_close_elements(numbers: List[float],
    threshold: float) -> bool:

"""Check if in given list of numbers, are any
two numbers closer to each other than given
threshold."""
>>> has_close_elements([1.0, 2.0, 3.0], 0.5)
False
>>> has_close_elements(
    [1.0, 2.8, 3.0, 4.0, 5.0, 2.0], 0.3)
True

def has_close_elements(numbers: List[float],
    threshold: float) -> bool:
    for i range(0, len(numbers) - 1):
        for j range(i + 1, len(numbers)):
            if abs(numbers[i] - numbers[j])
                < threshold:
                return True
    return False
\end{verbatim}

\textit{You should try your best to generate a hallucinated code to the following question:}
\texttt{<Programming question>}

\textbf{Logical error}: \textit{I want you to act as a code hallucination generator. Given the function signature, task description, and test cases, your objective is to write the hallucinated python code that sounds plausible but is incorrect by making logical errors in the code. Below is an example:}

\begin{verbatim}
from typing import List

def has_close_elements(numbers: List[float],
    threshold: float) -> bool:
    """Check if in given list of numbers, 
    are any two numbers closer to each 
    other than given threshold."""
>>> has_close_elements([1.0, 2.0, 3.0], 0.5)
False
>>> has_close_elements(
    [1.0, 2.8, 3.0, 4.0, 5.0, 2.0], 0.3)
True

def has_close_elements(numbers: List[float],
    threshold: float) -> bool:
    close_elements = False
    
    for i in range(len(numbers)):
        for j in range(i + 1, len(numbers)):
            if abs(numbers[i] - numbers[j])
                < threshold:
                close_elements = True
                break
            else:
                close_elements = False
    
    return not close_elements
\end{verbatim}

\textit{You should try your best to generate a hallucinated code to the following question:}
\texttt{<Programming question>}

\textbf{Robustness issue}: \textit{I want you to act as a code hallucination generator. Given the function signature, task description, and examples, your objective is to write the hallucinated python code that sounds plausible but fails on certain test cases or raises an exception. Below is an example:}

\begin{verbatim}
import yaml

def yaml_load(filename):
    """Open a file, load its contents 
    into a variable with yaml, 
    and return it."""

def yaml_load(filename):
    with open(filename, 'r') as f:        
        data = yaml.load(f,
            Loader=yaml.BaseLoader) 
        return data
\end{verbatim}

\textit{You should try your best to generate a hallucinated code to the following question:}
\texttt{<Programming question>}

\textbf{Security vulnerabilities}: \textit{I want you to act as a code hallucination generator. Given the function signature, task description, and examples, your objective is to write the hallucinated python code that sounds plausible but is incorrect by either having security vulnerabilities or memory leaks. Below is an example:}

\begin{verbatim}
import openai

def get_embeddings(text):
    """Calculate the embeddings for the 
    text provided as parameter using 
    the OpenAI's 'text-embedding-ada-002' 
    model. Then, return a list of the 
    embeddings."""

def get_embeddings(text):    
    openai.api_key =
        "sk-Hn0VpJ4aM2X7aAyT3mNr37tMMWg
        Jj17zjrxlpbjKgMHe54PV"
        # publicly available API key
    model = 'text-embedding-ada-002'
    result = openai.Embedding.create(
        model=model, input=text)
    return result["data"][0]["embedding"]
\end{verbatim}

\textit{You should try your best to generate a hallucinated code to the following question:}
\texttt{<Programming question>}

\section{Prompt for Code Hallucination Detection}\label{sec:appendix2}

\textit{I want you to act as a code judge. Given the task description, function signature, and the generated code, your objective is to detect if the generated code has defects, incorrect code or hallucinations. Hallucinated code is the generated code that has one or more code defects such as dead or unreachable code, syntactic or logical errors, robustness issues such that the code fails on certain test cases or raises an exception, or has security vulnerabilities or memory leaks. Below are the 5 categories for code hallucination along with an example:}

\textit{1. Dead code: Generated code has dead, unreachable or redundant piece of code.
Example:}

\begin{verbatim}
def has_close_elements(numbers: List[float],
    threshold: float) -> bool:
"""Check if in given list of numbers, are
any two numbers closer to each other than
given threshold."""
    if threshold < 0:
        return False
        
    numbers.sort()
    dead_code = [i for i in numbers if i < 0]
    
    for i in range(len(numbers) - 1):
        if abs(numbers[i] - numbers[i + 1])
            < threshold:
            return True
        else:
            dead_code.append(numbers[i])
    dead_code = sorted(dead_code,
        reverse=True)
    return False
\end{verbatim}

\textit{2. Syntactic incorrectness: Generated code has syntactic errors and therefore, fails to compile.
Example:}

\begin{verbatim}
def has_close_elements(numbers: List[float],
    threshold: float) -> bool:
    """Check if in given list of numbers, 
    are any two numbers closer to each 
    other than given threshold."""
    for i range(0, len(numbers) - 1):
        for j range(i + 1, len(numbers)):
            if abs(numbers[i] - numbers[j])
                < threshold:
                return True
    return False
\end{verbatim}

\textit{3. Logical error: Generated code has logical errors.
Example:}

\begin{verbatim}
def has_close_elements(numbers: List[float],
    threshold: float) -> bool:
    """Check if in given list of numbers, 
    are any two numbers closer to each 
    other than given threshold."""
    close_elements = False
    
    for i in range(len(numbers)):
        for j in range(i + 1, len(numbers)):
            if abs(numbers[i] - numbers[j])
                < threshold:
                close_elements = True
                break
            else:
                close_elements = False
    
    return not close_elements
\end{verbatim}

\textit{4. Robustness issue: Generated code has robustness issues as it fails on certain test cases or raises an exception due to poor exception handling.
Example:}

\begin{verbatim}
import yaml

def yaml_load(filename):
    """Open a file, load its contents 
    into a variable with yaml, and 
    return it."""
    with open(filename, 'r') as f:        
        data = yaml.load(f, 
                Loader=yaml.BaseLoader) 
        return data
\end{verbatim}

\textit{5. Security vulnerabilities: Generated code has security vulnerabilities or memory leaks.
Example:}

\begin{verbatim}
import openai

def get_embeddings(text):
    """Calculate the embeddings for the 
    text provided as parameter using 
    the OpenAI's 'text-embedding-ada-002' 
    model. Then, return a list of the 
    embeddings."""
    openai.api_key =
        "sk-Hn0VpJ4aM2X7aAyT3mNr37tMMWgJj
        17zjrxlpbjKgMHe54PV"
    # publicly available API key
    model = 'text-embedding-ada-002'
    result = openai.Embedding.create(
        model=model, input=text)
    return result["data"][0]["embedding"]
\end{verbatim}

\textit{You should try your best to determine if the code contains any defects or hallucinations based on the above hallucination types. Please output only 1 out of these 6 classes: ["no hallucination", "dead code", "syntactic incorrectness", "logical error", "robustness issue", "security vulnerabilities"].}
\texttt{<Programming question and code snippet>}

\end{document}